\documentclass[]{aa}
\usepackage{txfonts}
\usepackage{epsfig}
\usepackage{graphicx}
\usepackage{natbib}
\bibpunct{(}{)}{,}{a}{}{}

\def\4U{\object{MXB~1728--34}}
\begin{document}

\title{The Iron K-shell features of \4U from a Simultaneous 
Chandra-RXTE Observation}
  \subtitle{}
   \author{A. D' A{\'\i}\inst{1}, T. Di Salvo\inst{1}, R. Iaria\inst{1}, 
   M. M\'endez\inst{2}, L. Burderi\inst{3}, G. Lavagetto\inst{1},
   W. H.G. Lewin\inst{4}, N. R. Robba\inst{1}, L. Stella\inst{3}, 
   M. van der Klis\inst{5}}
   \offprints{A. D' A{\'\i}, dai@fisica.unipa.it}
   \institute{Dipartimento   di  Scienze   Fisiche   ed  Astronomiche,
              Universit\`a  di  Palermo,  via  Archirafi  36  -  90123
              Palermo,  Italy
              \and National Institute for Space Research, Sorbonnelaan
              2,  3584 CA Utrecht,  the Netherlands  
              \and  Osservatorio Astronomico  di Roma,
              via Frascati 33,  00040 Monteporzio Catone (Roma), Italy 
	      \and  Center for Space Research and  Department of Physics, 
              Massachusetts Institute of Technology, 70 Vassar Street, 
              Cambridge, MA 02139
	      \and Astronomical Institute  ``Anton Pannekoek",  University  
	      of Amsterdam and Center for  High-Energy Astrophysics, 
	      Kruislaan 403, NL 1098  SJ Amsterdam,  the Netherlands }
\date{Received / Accepted}
\authorrunning{A. D' A{\'\i} et al.} 
\titlerunning  {The iron K-shell features of \4U}
\abstract{
  We report on a simultaneous Chandra and RossiXTE observation of the
  low-mass X-ray  binary atoll bursting  source \4U performed  on 2002
  March  3--5.  We  fit  the  1.2--35 keV  continuum  spectrum with  a
  blackbody plus  a Comptonized  component.  Large residuals  at 6--10
  keV can be fitted by a broad (FWHM $\simeq 2$ keV) Gaussian emission
  line  or, alternatively,  by  two absorption  edges associated  with
  lowly ionized  iron and Fe XXV/XXVI  at $\sim 7.1$ keV  and $\sim 9$
  keV, respectively.   In this interpretation, we find  no evidence of
  broad, or narrow,  emission lines between 6 and 7  keV.  We test our
  alternative modelling  of the iron  K shell region by  reanalysing a
  previous BeppoSAX  observation of  \4U, finding a  general agreement
  with our  new spectral model.   
}
\maketitle
\section{Introduction}
\4U (discovered by \citealt{lewin76}; see also \citealt{hoffman76}) is
a low-mass  X-ray binary, a well  known prototype of the  class of the
bursting atoll sources \citep{hasinger89}.  The distance to the source
is   estimated  between   4.1  kpc   and  5.1   kpc  \citep{disalvo00,
  galloway03}.  This was one of the first sources to display kilohertz
quasi-periodic oscillations  (kHz QPOs) in its power  spectrum and the
first to  display quasi-coherent  oscillations, now identified  as the
spin frequency  of the neutron star  (NS), around 363  Hz, during some
type I X-ray bursts  \citep{strohmayer96}.  Its temporal behaviour has
been     extensively      studied     by     \citet{disalvo01}     and
\citet{vanstraaten02},  using  a   large  set  of  RXTE  observations,
spanning more than three years.

Power spectra of NS systems  show a variety of rapid X-ray variability
components  and noise  features.  Among  the  high-frequency phenomena
kilohertz  quasi-periodic  oscillations  (kHz  QPOs) are  the  fastest
variability  components.  Two  QPO peaks  usually occur  in  the range
200--1200 Hz, the higher frequency  QPO denoted the upper kHz QPO, the
other the  lower kHz QPO; their  frequencies move to  higher values as
the spectral  state of the  source changes, while the  peak separation
remains almost constant, probably  being related to the spin frequency
of the neutron  star or to half its  value \citep[see e.g.][]{klis00}.
Neutron-star  hectohertz  QPO  is  a  peaked noise  component  in  the
frequency  range  100--200  Hz,  which  does  not  present  any  clear
correlation with  other features  in the power  spectrum but  could be
linked to the high-frequency  QPOs observed in some black-hole systems
\citep{nowak00}.  The  low frequency complex  is dominated by  a broad
band   noise  component,   usually  modelled   with   a  zero-centered
Lorentzian, and one or two broad QPOs in the range 0.1--50 Hz, usually
denoted Low-Frequency QPO (LFQPO) and Very Low Frequency QPO (VLFQPO).

Spectral studies  have been  carried out in  the past with  the EXOSAT
\citep{white86}, RXTE \citep{piraino00} and BeppoSAX \citep{disalvo00,
  piraino00}  satellites.  \citet{white86}  found  that the  continuum
emission  could  be well  described  by  a  Comptonization model,  and
interpreted this result in the framework of the so called {\it Western
  Model}, where  the majority of  the X-ray emission is  disk emission
Comptonized in a sorrounding corona (but in bright sources a blackbody
component,  interpreted  as emission  from  the  NS  surface was  also
modeled,    see   also~\citealt{white88}).     \citet{disalvo00}   and
\citet{piraino00} found that a two-component model was required to fit
the  broad energy  band spectra  of  \4U from  BeppoSAX (0.1--200  keV
range)  and RXTE  (3.0--60 keV  range).  The  spectra, in  both cases,
consisted  of a  soft  component  (a multicolor  disk  blackbody or  a
blackbody spectrum), probably produced  by the disc, and a Comptonized
spectrum,  responsible  for  the  hard emission,  where  seed  photons
arising from  the neutron  star, or from  an optically  thick boundary
layer, are  scattered by a  small spherical corona placed  between the
innermost edge of  the accretion disc and the  neutron star surface (a
description  known   in  literature   as  the  {\it   Eastern  Model},
\citealt{mitsuda89}).  More  recently, detailed and  extensive studies
on   dipping    X-ray   sources   \citep{balucinska99,   balucinska00,
  balucinska04} have shown that the  size of the accretion disk corona
(ADC, $\simeq$ 50,000  km) largely exceeds the inferred  inner edge of
the  accretion disk,  so that  the majority  of the  disk  emission is
expected to be strongly Comptonized.  The {\it Birmingham Model}, i.e.
a blackbody emission  from the NS plus a  Comptonized emission from an
extended accretion  disk corona, therefore, seems  now physically more
appropriate in describing the continuum emission from LMXBs \citep[see
also][] {church01}.  However, different interpretations of the dipping
behaviour  and the  consequent continuum  modelling are  still debated
\citep{boirin05}.

Part of  the hard X-ray emission  could evenually be  scattered by the
disc, producing  Compton reflection features such as  a broad Gaussian
line at  6.4--6.7 keV.  However, the  width of the  line inferred from
the fits ($\geq$ 0.5 keV) lacks a clear physical interpretation, as up
to date  it is yet disputed  how this width could  actually arise (see
also \citealt{asai00} for a critical discussion on possible broadening
mechanisms).

In this paper we present  new results from a simultaneous Chandra-RXTE
observation of \4U. We propose  an alternative fit of the iron K-shell
region, using  two absorption edges  instead of a very  broad emission
line, the interpretation of which could be quite problematic.

\section{Observations and Data Reduction}

\4U was observed with Chandra from 2002 March 4 15:20:34 to 2002 March
5  00:27:26 using  the High  Energy Transmission  Grating Spectrometer
(HETGS).  The data were collected  in the Timed Exposure Mode, using a
subarray of  the ACIS-S  cameras in order  to mitigate the  effects of
photon  pile-up in the  first order  spectra.  Consequently  the frame
time was  1.44 s, the High  Energy Grating (HEG)  spectrum was cut-off
below 1.6 keV  and the Medium Energy Grating  (MEG) spectrum below 1.2
keV. The RXTE  observation started on 2002 March  3 03:27:12 and ended
on 2002 March 5 13:00:00.  We used only the Proportional Counter Array
(PCA) data;  the total available time,  i.e. the time  during which at
least one Proportional  Counter Unity (PCU) was on,  was $\sim 97$ ks.
For the timing analysis we used only PCU0 and PCU2, which are the PCUs
with the longest simultaneous on-source time (almost 90 ks), while for
the spectral analysis we used PCU2  and PCU3, avoiding the use of PCU0
because starting 2000 May 12 it  suffered the loss of the veto propane
layer.  We used Standard2 configuration, with 16 s time resolution and
129  energy  channels,  for  the  spectral  analysis  and  Event  Mode
configuration,  with  128  $\mu  s$  time  resolution  and  64  energy
channels, for the temporal analysis.

We used the CIAO 3.0 software package for the reduction of the Chandra
data, FTOOLS 5.3  for the reduction and analysis of  the PCA data, and
XSPEC v. 11.3 for the spectral analysis.  Because the Rapid Burster is
inside the field of view of  the PCA ($\sim 1^{\circ}$) we checked the
flux  of this  source during  our  observation analyzing  the All  Sky
Monitor (ASM) lightcurve.   We found that the source  did not show any
significant activity so  that the associated flux should  be less than
$1.4  \times  10^{-11}$  erg cm$^{-2}$  sec$^{-1}$  \citep{masetti00}.
Because  the  flux  of  \4U  in  the  same  energy  range  during  our
observation   was  $\sim$   1.5  $\times$   10$^{-9}$   erg  cm$^{-2}$
sec$^{-1}$,  we  concluded that  the  Rapid  Burster contamination  is
absolutely negligible.\\  In order  to constrain the  systematic error
for  the PCA spectra  we analyzed  an observation  of the  Crab Nebula
(Obs.  ID  70018-01-01-00), performed 50 min after  our observation of
\4U.   We fitted  the Crab  spectrum with  a  simple photoelectrically
absorbed  power-law model and  derived the  systematic error  from the
residuals  in the fit  as described  in \citet{wilms99},  treating the
spectra  of different PCUs  as independent  datasets fitted  under the
same  model.  We  raised the  level of  systematics until  a  value of
$\chi^2~<~2$  was  obtained.   These  operations  resulted  in  a  2\%
systematic  error  for  channels  below  25 keV,  and  2.5\%  for  the
remaining channels.

Seven type-I bursts were revealed in the PCA lightcurve and two bursts
in the  Chandra lightcurve.   As our primary  concern is  studying the
persistent emission of the source we discarded data intervals centered
around each burst, for a time  length of 160 sec.  The persistent flux
of   the  source,   during  the   whole  observation,   showed  little
variability,  and  remained at  about  20--25  cts/s  for the  Chandra
diffraction arms  and 235--270 cts/s/PCU for the  PCA, thus confirming
that any contamination of the Rapid Burster can be safely neglected.

\section{Temporal Analysis}

We studied the Power Density Spectrum (PDS) of \4U using the high-time
resolution data  of the RXTE/PCA.  The power  spectra were constructed
by  dividing  the  PCA  lightcurve,  with no  energy  selection,  into
segments of  256 s and  then binning the  data in time  before Fourier
transforming such that the Nyquist frequency was 2048 Hz in all cases;
the  PDS were normalized  according to  \citet{leahy83}.  All  the PDS
produced  were finally  summed and  averaged into  a single  PDS.  The
Poisson noise estimated  in the frequency range between  1200 and 2048
Hz  (where  neither  noise nor  QPOs  are  known  to be  present)  was
subtracted before fitting.

A satisfactory  fit of  the PDS is  obtained by decomposing  the power
spectrum into a sum of five Lorentzians \citep[see e.g.][]{belloni02}:
a zero-centered Lorentzian, which  accounts for the band-limited noise
component (BLN  Lorentzian), a Lorentzian  centered at $\simeq$  23 Hz
(the VLFQPO Lorentzian, L$_{VLF}$),  a Lorentzian centered at $\simeq$
38 Hz (the LFQPO Lorentzian, L$_{LF}$), a broad Lorentzian centered at
$\simeq$  130  Hz  (the  hectohertz  QPO, L$_{hHz}$),  and  finally  a
Lorentzian at  $\simeq$ 650 Hz  (the kHz QPO,  L$_{kHz}$).  Parameters
best-fit  values   and  associated  errors,   calculated  for  $\Delta
\chi^2=1.0$, equivalent  to 1 $\sigma$ confidence level,  for this fit
are reported in Table \ref{table1}.

It has been shown that the  value of the centroid frequency of the kHz
QPOs  correlates  well  with  the   position  of  the  source  in  the
color-color  diagram (CD),  gradually increasing  as the  source moves
from the Island State to the Banana State \citep[e.g.][]{klis00}.  The
average PDS during our observations  looks similar to the PDS numbered
7 in \citet{disalvo01}  when the source lies in the  lower part of its
Island State.  The  high value of the rms  fractional amplitude of the
L$_{kHz}$  detected  during   the  Chandra-RXTE  observation  strongly
suggests it  is the upper  peak of the  twin kHz QPOs;  following this
identification, we  find that the  parameters of the  other components
obtained  from our  fit are  in good  agreement with  the correlations
shown  in \citet{disalvo01},  except that  we detect  L$_{VLF}$, which
\citet{vanstraaten02} argued  could only appear as both  the lower and
the upper kHzQPOs  were present.  The presence of  this feature during
our observation may be  linked to the frequency statistical resolution
of  the power  spectra, as  our integration  time greatly  exceeds the
integration  time  used  in  past analyses.   Integrating  on  shorter
timescales ($\leq$~20  ksec) we  found that the  VLFQPO was  no longer
required  for  a good  fit.  In  this case,  the  best  value of  peak
frequency  of the LFQPO  stayed almost  constant at  23 Hz,  while the
value  of the  parameters  of  the other  Lorentzians  did not  change
significantly, so  that our analysis  would coherently agree  with the
previous ones.

Both the  timing properties and the  position of the source  in the CD
track suggest that during our observation \4U was in the Island State,
corresponding  to a  relatively low,  or intermediate,  mass accretion
rate.
\begin{table} 
\caption{ \footnotesize \linespread{1} 
Results of the \4U PDS fit in the frequency range $3.9 \times 10^{-3} -
2048$ Hz. Here $\nu_0$ is the
centroid frequency and $\Delta$ is  full width at half maximum of each
Lorentzian line.   The  root-mean-square   amplitude  (RMS)  is  calculated
integrating each  Lorentzian from 0  to infinity.  Errors are  quoted 
at $1 \sigma$ confidence level.}
\label{table1}
\centering
\linespread{1}
\begin{tabular}{l l l}
\hline
\hline
Component           &      Parameter                & Value \\
\hline
BLN      & $\nu_0$  (Hz)           & 0 $(fixed)$\\
BLN      & $\Delta$ (Hz)  & 20.9$\pm$0.4 \\
BLN      & RMS (\%)       & 21.6$\pm$0.5 \\
\hline
L$_{VLF}$   & $\nu_0$  (Hz)                & 23.2$\pm$0.3\\
L$_{VLF}$   & $\Delta$ (Hz)                & 11.5$\pm$1.5\\
L$_{VLF}$   & RMS (\%)                     & 6.0$\pm$0.8 \\
\hline
L$_{LF}$      & $\nu_0$  (Hz)  & 38.1$\pm$1.7 \\
L$_{LF}$      & $\Delta$ (Hz)   & 26.8$\pm$2.0 \\
L$_{LF}$    & RMS (\%)        & 7.0$\pm$0.7\\
\hline
L$_{hHz}$ & $\nu_0$  (Hz)     & 134$\pm$7 \\
L$_{hHz}$ & $\Delta$ (Hz)    & 221$\pm$22 \\
L$_{hHz}$ & RMS (\%)         & 12.4$\pm$0.7\\
\hline
L$_{kHz}$     & $\nu_0$  (Hz)     & 647$\pm$5 \\
L$_{kHz}$     & $\Delta$ (Hz)      & 175$\pm$8 \\
L$_{kHz}$     & RMS (\%)        & 10.4$\pm$0.7 \\
\hline
$\chi^2$/dof &                  & 512/432 \\
\hline
\hline
\end{tabular}
\end{table}
\section{Spectral Analysis}
For  the  Chandra/HETG  data,   we  considered  the  four  first-order
dispersed spectra, namely the two  HEG spectra and the two MEG spectra
on  opposite sides  of  the  zeroth order  image.   We averaged  HEG+1
(MEG+1) and HEG-1 (MEG-1) spectra into a single spectrum, after having
tested their reciprocal consistency.  We  used data in the 1.6--10 keV
range for the HEG spectrum and  1.2--5 keV range for the MEG spectrum. 
For all the fits we took  into account an instrumental feature at 2.07
keV for bright sources, described  by \citet{miller02}, and fit it with
an inverse edge (with optical  depth $\tau \simeq -0.1$).  The HEG and
MEG spectra were binned in order  to have at least 300 counts for each
bin.  This,  however, still ensures  a high number of  channels (about
1000) and  good  spectral  resolution  throughout the  entire  covered
energy band.

For the RXTE/PCA  data we applied the standard  selection criteria for
obtaining good time intervals. We  limited the analysis to the 3.5--35
keV energy range and applied a systematic error as described in \S 2.

We restricted the spectral analysis to the intervals during which RXTE
operated simultaneously with  Chandra.  Relative normalizations of the
three instruments, except for HEG which was fixed to a reference value
of 1, were left as free parameters in all the fits.

We applied a series of models to simultaneously fit HEG, MEG, and RXTE
spectra.   Because in this  paper we  are interested  in the  study of
discrete features, we used to describe the continuum the two-component
model which gave  the lowest value of $\chi^2$.   We find the best-fit
continuum  model  to consist  of  a  soft  component, described  by  a
blackbody of temperature $kT_{bb}~\simeq$ 0.51 keV, plus a Comptonized
component  (CompTT in  XSPEC;  \citet{titarchuk94}), with  seed-photon
temperature $kT_0~\simeq$ 1.3  keV, electron temperature $kT_e~\simeq$
7.4  keV  and  optical   depth,  associated  to  a  spherical  corona,
$\tau~\simeq$ 6.2.  Both  components are photoelectrically absorbed by
an    equivalent   hydrogen    column   $N_H    \simeq    2.6   \times
10^{22}$~cm$^{-2}$.   The  $\chi^2$/dof   obtained  for  this  fit  is
1206/1034.

An  absorption feature,  resembling  an edge,  at  1.84 keV,  probably
associated  to neutral  Si,  is clearly  visible  in the  MEG and  HEG
spectra (see Fig.~1, left panel).  To fit this edge we substituted the
absorption  component {\it  phabs} in  XSPEC with  the  component {\it
  vphabs}, which allows  us to vary the abundances  of single elements
with respect  to the  solar abundances.  We  find that a  Si abundance
$\sim$  2 times  larger  than  the solar  abundance  improves the  fit
significantly ($\chi^2$/dof = 1031/1033 for this fit).

With respect to any continuum  model that we tried, large residuals in
the 6--10 keV range are clearly  visible (Fig.  1, left panel) and can
be  fitted  by a  very  broad  Gaussian  emission line.   Leaving  the
parameters  of  the  line  free  gives,  however,  unphysical  results
(position of the  line $\simeq$ 5.5 keV and  $\sigma \simeq$ 1.5 keV).
Confining the position  of the line in the range  6.4--6.7 keV gives a
line width $\sigma \simeq 1$  keV (equivalent width $\simeq 270 $ eV);
fixing the  width of the  line to 0.5  keV constrains its  position at
$\simeq 6.1$ keV. Alternatively, we  can also fit these residuals with
absorption  edges, instead  of emission  lines, associated  with lowly
ionized iron  and Fe XXV/XXVI at  $7.03$ keV ($\tau  \simeq 0.11$) and
$\sim  9$ keV  ($\tau \simeq  0.16$), respectively.   The  addition of
these  edges improves  the fit  significantly compared  to  the simple
blackbody   plus   Comptonization   model  described   above,   giving
$\chi^2$/dof of  959/1029.  In this interpretation, we  no longer find
evidence of broad or narrow iron emission lines between 6 and 7 keV.

To  test  our model  we  reanalyzed  a  previous BeppoSAX  observation
performed  between August  23 and  24,  1998 \citep[see][]{disalvo00}.
During this observation the ASM lightcurve of the Rapid Burster showed
that  the source  was in  an  active state  (10.7/11.8 cts/s,  one-day
average). LECS  and MECS spectra, which were  extracted using circular
regions around \4U  of 8' and 4' radii  respectively, are not affected
by the  contaminating source.  HP and PDS  spectra , extracted  from a
1$^\circ$ region, can  be affected up to 10-15\%  for the source flux.
The normalization constants among the MECS/LECS/PDS/HP instruments are
inside  the expected calibration  ranges so  that, although  we cannot
exclude some contamination  in the residuls of our  fits above 10 keV,
our conclusions in the low energy band of the spectrum still hold.

We fittted the  0.12-60 keV BeppoSAX spectrum with  the same continuum
adopted for the  Chandra-RXTE dataset. A Gaussian emission  line is no
longer required if we introduce two absorption edges at energies above
7  keV.  In this  case, we  find the  first edge  at $\simeq$  7.5 keV
($\tau$ $\simeq$ 0.08) and the second edge at $\simeq$ 8.8 keV ($\tau$
$\simeq$  0.08).   The BeppoSAX  spectrum  is  in  agreement with  the
Chandra-RXTE  spectrum, giving  a $\chi^2$/dof  = 207/178,  instead of
$\chi^2$/dof   =   235/180  obtained   for   the   model  adopted   in
\citet{disalvo00}).  Note that the overabundance of Si with respect to
Solar  abundance required  to  fit the  Chandra-RXTE  spectrum is  not
required  to  fit the  BeppoSAX  spectrum.   On  the other  hand,  the
BeppoSAX  spectrum  shows  in  this  energy band  a  prominent  narrow
emission line at  1.67 keV (Fig.  2, left  panel), compatible with the
radiative recombination emission  from Mg XI \citep[see][]{disalvo00}.
This  feature,  however, is  not  detected  in  the Chandra  spectrum.
Removing this  emission line from  the fit gives a  worse $\chi^2$/dof
but  a Si-overabundance  is required  to  fit the  data, resulting  in
values also in  agreement with the Chandra-RXTE dataset;  in this case
we find that  $\chi^2$/dof is 225/179 when the Si  abundance is $2 \pm
0.5$ times solar, and $\chi^2$/dof is 247/180 when the Si abundance is
fixed to the solar value.

\begin{figure*}
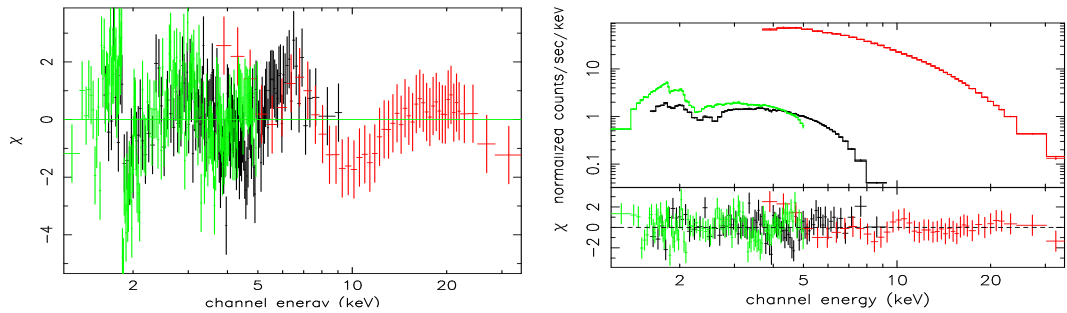

\centering
\caption{  \footnotesize \linespread{1}  
  1.2--35.0 keV spectra of  \4U obtained from the simultaneous Chandra
  and RXTE datasets.   Left panel: residuals in unit  of $\sigma$ with
  respect  to the absorbed  blackbody plus  CompTT model.  Right upper
  panel:  Chandra and RXTE  /PCA spectra  together with  the two-edges
  model (see  table 2) shown on top  of the data as  solid line; right
  lower panel: residuals to  the best-fit  model.  Data
  have  been rebinned  for clarity:  MEG data  in green,  HEG  data in
  black, PCA data in red.}
\label{spettro_chandra}
\begin{tabular}{l l}
\includegraphics[height=6.8cm, width=4cm, angle=-90]{residuals_chandraXTE.cps} & 
\includegraphics[height=6.8cm, width=4cm, angle=-90]{spettro_chandra_edge.cps}     \\
\end{tabular}
\end{figure*}

\begin{figure*}
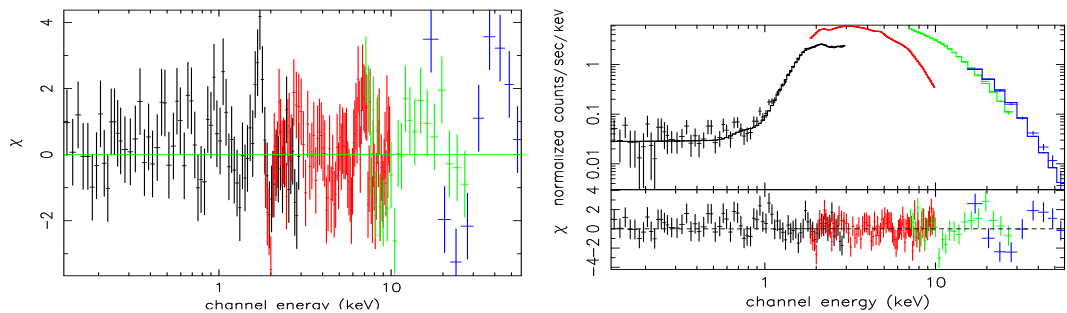

\centering
\caption{  \footnotesize \linespread{1}  
  0.12--60 keV spectra of  \4U obtained from the BeppoSAX observation.
  Left  panel: residuals  in  unit  of $\sigma$  with  respect to  the
  absorbed blackbody  plus CompTT model.  Right  upper panel: BeppoSAX
  spectra together with the two-edges model (see table 2) shown on top
  of  the data  as solid  line; right  lower panel:  residuals  to the
  best-fit model.}
\label{spettro_sax}
\begin{tabular}{l l}
\includegraphics[height=6.8cm, width=4cm, angle=-90]{residuisax.cps} & 
\includegraphics[height=6.8cm, width=4cm, angle=-90]{spettro_sax_edge.cps}     \\
\end{tabular}
\end{figure*}

The best-fit parameters for the  fits of the Chandra-RXTE and BeppoSAX
datasets are reported in  \ref{table2}; residuals in units of $\sigma$
with   respect  to   the   simple  continuum   model   are  shown   in
Figures~\ref{spettro_chandra}  and  \ref{spettro_sax}  (left  panels);
data and  residuals corresponding to the  two edge model  are shown in
Figures~\ref{spettro_chandra} and~\ref{spettro_sax} (right panels).
\begin{table*}
\centering
\caption{ \footnotesize  \linespread{1}
  Best fit parameters of \4U for the Chandra/RXTE spectra (1.2--35
  keV) and  the BeppoSAX spectra (0.12--60 keV  energy band).  The
  continuum emission consists of a blackbody (bbody) and a Comptonized
  component   (compTT).    $kT_{\rm  BB}$   and   N$_{\rm  BB}$   are,
  respectively, the  blackbody temperature and  normalization in units
  of $L_{39}/D_{10}^2$,  where $L_{39}$ is the luminosity  in units of
  $10^{39}$ ergs/s  and $D_{10}$ is the  distance in units  of 10 kpc.
  $kT_0$,   $kT_{\rm   e}$  and   $\tau$   indicate  the   seed-photon
  temperature, the electron temperature,  and the optical depth of the
  spherical Comptonizing cloud.  N$_{\rm CompTT}$ is the normalization
  of the CompTT model  in XSPEC v.11.3 units.  Unabsorbed luminosities
  of the  bbody component and  of the CompTT component  are calculated
  assuming a distance  to the source of 5.1  kpc \citep{disalvo00} in
  the 0.1--100 keV energy  range.  For the edge components, E$_{edge}$
  denotes the energy of the edge and $\tau$ the optical depth. For the
  Gaussian component, E$_{Line}$ is the line centroid energy, $\sigma$
  the line  width, and N$_{\rm  Gauss}$ the total  photons/cm$^2$/s in
  the line.  The position  of the line at 1.66 keV had  to be fixed in
  order to make stable the  fit.  Uncertainties are given at the
  90\% confidence level.}
\label{table2}
\begin{tabular}{l l  |c|c|c|c}
  \hline
  \hline
  &                                  & Chandra-RXTE &  Chandra-RXTE & BeppoSAX               &  BeppoSAX   \\
  &                                  & 2 Edges      &  Broad Line   &  2 Edges               &  Broad Line \\
  \hline
  Component & Parameter (Units) & \multicolumn{4}{c}{Values} \\
  \hline

  vpha  & $N_{\rm  H}$ $(10^{22}$ cm$^{-2}) \dotfill$   & $2.61^{+0.06}_{-0.07}$  & $2.66^{+0.07}_{-0.07}$ & $2.31^{+0.08}_{-0.07}$  & $2.36^{+0.09}_{-0.08}$   \\
  vpha  & Si (Solar units)$\dotfill$                   & $2.02_{-0.13}^{+0.13}$  & $1.95_{-0.12}^{+0.11}$ &$\dots$ & $\dots$ \\

  edge  & E$_{edge}$(keV)  $\dotfill$    & $7.03^{+0.08}_{-0.06}$ & $\dots$  & $7.47^{+0.14}_{-0.13}$    & $\dots$\\
  edge  &  $\tau$ $(10^{-2})$ $\dotfill$ & $11^{+3}_{-4}$         &  $\dots$ & $8^{+2}_{-2}$     & $\dots$  \\

  edge  & E$_{edge}$ (keV)   $\dotfill$   & $9.0^{+0.3}_{-0.4}$ & $\dots$ & $8.82^{+0.2}_{-0.19}$   & $\dots$\\
  edge  & $\tau$ $(10^{-2})$ $\dotfill$   & $16^{+4}_{-4}$      & $\dots$ & $8.82^{+0.2}_{-0.19}$   & $\dots$ \\

 Gauss & E$_{Line}$ (keV)      $\dotfill$        &      $\dots$               & $6.12^{+0.19}_{-0.19}$ & $\dots$              & $6.74^{+0.15}_{-0.15}$ \\
 Gauss & $\sigma$ (keV)           $\dotfill$     &       $\dots$              & $0.5$ (fixed) & $\dots$              & $0.5$ (fixed)\\ 
 Gauss & N$_{\rm  Gauss}$ $(10^{-3})$ $\dotfill$ &       $\dots$              & $2.3^{+0.5}_{-0.5}$& $\dots$              &  $1.7^{+0.5}_{-0.5}$\\

Gauss & E$_{Line}$ (keV)      $\dotfill$              &$\dots$ &$\dots$&$1.66$ (fixed)          & $1.66$ (fixed)\\
Gauss & $\sigma$ (keV)           $\dotfill$           &$\dots$ &$\dots$&$0^{+0.05}$             & $0^{+0.05}$    \\ 
Gauss & N$_{\rm  Gauss}$ $(10^{-3})$ $\dotfill$       &$\dots$ &$\dots$&$11^{+3}_{-3}$          & $11^{+0.3}_{-0.3}$   \\

  bbody & kT     (keV)   $\dotfill$                                & $0.513_{-0.014}^{+0.014}$ & $0.478_{-0.014}^{+0.014}$& $0.619_{-0.013}^{+0.013}$ & $0.590_{-0.017}^{+0.017}$\\  
  bbody & N$_{\rm  BB}$ $(10^{-3})$   $\dotfill$                   & $10.7_{-0.4}^{+0.4}$      & $10.1_{-0.5}^{+0.5}$& $19_{-5}^{+5}$            &$18_{-0.5}^{+0.6}$  \\ 
  bbody & L $(10^{36}$ erg sec$^{-1})$    $\dotfill$       & 2.78                    & 2.60 & 4.96                  & 4.69\\

  CompTT   & $kT_0$ (keV)         $\dotfill$                           & $1.33^{+0.05}_{-0.05}$ & $1.13^{+0.04}_{-0.04}$ & $1.62^{+0.03}_{-0.05}$ &$1.42^{+0.04}_{-0.04}$\\   
  CompTT   & $kT_{\rm  e}$ (keV)   $\dotfill$                          & $7.4_{-0.4}^{+0.5}$    & $7.2_{-0.5}^{+0.6}$ & $7.3_{-0.9}^{+1.5}$    & $5.3_{-0.3}^{+0.4}$ \\  
  CompTT   & $\tau$                  $\dotfill$                        & $6.2^{+0.4}_{-0.6}$    & $6.7^{+0.4}_{-0.4}$ & $6.2^{+0.4}_{-0.6}$    & $6.7^{+0.4}_{-0.4}$\\  
  CompTT   & N$_{\rm  CompTT}$ $(10^{-2})$     $\dotfill$              & $4.4^{+0.3}_{-0.6}$    & $4.7^{+0.4}_{-0.4}$& $5.6^{+1}_{-1.1}$    & $8.4^{+0.8}_{-0.7}$\\
  CompTT   & L $(10^{36}$ erg sec$^{-1})$$\dotfill$            & 9.0                      & 9.0 & 10.1                 &  10.1   \\ 
  \hline
  \multicolumn{2}{l}{$\chi^2$/dof} & 959/1029 & 978/1031 & 207/178 & 235/180 \\
  \hline
  \hline
\end{tabular}
\end{table*}
\section{Discussion}
Using the  HETGS on Chandra and the  PCA on RXTE we  obtained, for the
first time, simultaneous high-resolution and broad-band spectra of the
bursting X-ray  binary \4U.  The  source shows a quite  hard spectrum,
where  the  unabsorbed 0.1--10  keV  flux,  $2.5  \times 10^{-9}$  erg
cm$^{-2}$  sec$^{-1}$ is comparable  with the  10--200 keV  flux ($1.3
\times 10^{-9}$  erg cm$^{-2}$ sec$^{-1}$).   To fit the  continuum we
used the two-component model that  gave the minimum value of $\chi^2$,
which  consisted of blackbody  emission (the  radius of  the blackbody
emitting region is  $\sim 17.4$ km, for a distance of  5.1 kpc), and a
Comptonized  component.  Although  other  continuum models  have  been
proposed for the spectra of Low  Mass X-ray Binaries, the data of this
observation are not able  to discriminate among the proposed scenarios
as  slightly diffent values  of the  $\chi^2$ alone  do not  prove one
model  is corret  \citep{schulz93}. Consequently  we will  not further
discuss the physical implications  of the parameter values obtained by
our best fit, but we will  concentrate on the the complex structure in
the 6--9 keV range, where several Fe K lines and edges are expected to
be  present.  We  found  that in  the  high-resolution spectrum  these
features are  present independently of  the continuum model.   Fits to
broad-band low-resolution BeppoSAX spectra  of this source in the past
included  a  broad  Gaussian line  at  $\simeq$  6.7  keV to  fit  the
structure in the 6--9 keV range.  We find that a model consisting of a
combination  of absorption features  due to  lowly and  highly ionized
iron  on top  of a  broad-band  continuum fits  the Chandra-RXTE  data
significantly better.  Furthermore, the same absorption model fits the
BeppoSAX data better than the one with the broad emission line.

In this interpretation we report a shift in the energy of the first of
these edges, from 7.03 keV during the Chandra-RXTE observation to 7.46
keV (compatible with K$\alpha$ edges of moderately ionized iron, Fe IX
to Fe  XVI) during the BeppoSAX  observation, while the  energy of the
second edge, interpreted as due  to highly ionized (Fe XXV/XXVI) iron,
is consistent with  being the same in both  observations.  This is the
first  direct  evidence of  the  existence  of photoionized  absorbing
material in the  vicinity of \4U. The difference in  the energy of the
moderately ionised  iron edge can be  due to the fact  that during the
BeppoSAX  observation the  source  was more  luminous  (see Table  1),
reflecting a higher accretion rate.  The absorbing regions responsible
for the two edges should  then be spatially separated.  The first edge
(implying low ionization states)  may originate from absorption of the
primary continuous  emission in the  external regions of  an accretion
disk \citep[see also][]{singh94}, while the second edge (implying much
higher ionization  states) may be  produced in the  hot (photoionized)
Comptonizing corona.   \citet{boirin04} reported  a list of  known low
mass  X-ray binary systems  that exhibit  in their  spectra absorption
features  related to  the presence  of highly  ionized iron  and other
metals.  Almost all these systems  are seen at high inclination angles
($i > 60^o$), i.e.  almost edge  on.  We presume that \4U, although it
never displayed  a dipping behaviour, could  also be a  member of this
class  of  objects  and   consequently  be  a  system  at  medium-high
inclination angle.

An overabundance of neutral Si by a factor of $\sim 2$ with respect to
the solar  abundance is required to  fit the HEG and  MEG spectra, but
not necessarily  to fit a  non-simultaneous BeppoSAX spectrum  of this
source.   We   presume  that  this  feature  may,   therefore,  be  of
instrumental origin.
\begin{acknowledgements}
  This work was partially supported by the Ministero della Istruzione, 
della Universit\'a e della Ricerca (MIUR).
\end{acknowledgements}
\bibliographystyle{aa}
\bibliography{references}
\end{document}